\documentclass[letterpaper,notoc]{JHEP3}

\usepackage{epsfig,multicol}

\newcommand\fverb{\setbox\fverbbox=\hbox\bgroup\verb}
\newcommand\fverbdo{\egroup\medskip\noindent\fbox{\unhbox\fverbbox}\ }
\newcommand\fverbit{\egroup\item[\fbox{\unhbox\fverbbox}]}
\newbox\fverbbox

\title{The influence of collective neutrino oscillations on a supernova $r$ process}
\author{Huaiyu Duan \\ Department of Physics and Astronomy, University of New Mexico, Albuquerque, NM 87131, USA}
\author{Alexander Friedland \\ Theoretical Division, Los Alamos National Laboratory, Los Alamos, NM 87544, USA}
\author{Gail C. McLaughlin \\ Department of Physics, NC State University, Raleigh, NC 27695-8202, USA}
\author{Rebecca Surman \\ Department of Physics and Astronomy, Union College, Schenectady, NY 12308, USA}

\abstract{Recently, it has been demonstrated that neutrinos in a supernova oscillate
\emph{collectively}. This process occurs much deeper than the conventional matter-induced MSW effect
and hence may have an impact on nucleosynthesis.  In this paper we explore the effects of collective
neutrino oscillations on the $r$-process, using representative late-time neutrino spectra and outflow
models. We find that accurate modeling of the collective oscillations is essential for this analysis.
As an illustration, the often-used ``single-angle'' approximation makes grossly inaccurate predictions
for the yields in our setup. With the proper multiangle treatment, the effect of the oscillations is
found to be less dramatic, but still significant. Since the oscillation patterns are sensitive to the
details of the emitted fluxes and the sign of the neutrino mass hierarchy, so are the $r$-process
yields. The magnitude of the effect also depends sensitively on the astrophysical conditions --- in
particular on the interplay between the time when nuclei begin to exist in significant numbers and the
time when the collective oscillation begins.  A more definitive understanding of the astrophysical
conditions, and accurate modeling of the collective oscillations for those conditions, is necessary. }

\preprint{LA-UR-10-06628\\INT-PUB-10-065}

\begin{document}

\section{Introduction}

 There are a number of types of element synthesis that are thought to occur in ``hot outflows'', such as the formation
of light $p$-process elements, possibly $r$-process elements, and also some iron peak elements.  A hot outflow is one in
which nucleons are originally photo-dissociated into free nucleons at temperatures of around a few tens of MeV. 
Material typically flows away from a hot center, such as an accretion disk around a black hole or proto-neutron star,
and as it does so the nucleons cool and combine into nuclei.  Given the high temperature of the center, neutrinos can be
emitted in vast numbers. These neutrinos interact with the outflowing and cooling nucleons and nuclei and, among other
things, change neutrons into protons and vice-versa.  Thus any process which alters the luminosity or spectra of the
neutrinos may impact element synthesis. Neutrino flavor transformations -- and, specifically, collective oscillations,
as explained later -- are a prime example of such a process and it is essential to determine whether they are important
for nucleosynthesis in hot outflow environments.  In this initial work we explore, by considering a few specific
examples, the size of this effect on the abundances of rapid neutron capture elements.

It should be stressed that the physics underlying the oscillations is by now well established and treating it is no
longer optional. The essential ingredients are the measured neutrino oscillation parameters and, crucially, the
neutrino-neutrino coherent interactions
\cite{Fuller:1987,Notzold:1987ik,Pantaleone:1992xh,Pantaleone:1992eq,Sigl:1992fn,McKellar:1992ja}, which can cause
\emph{collective} oscillations. It has been suspected for nearly two decades that neutrino oscillations ({\it e.g.},
\cite{Qian:1993dg}), particularly the collective oscillations between the active flavors
\cite{Pantaleone:1994ns,Qian:1994wh,Sigl:1994hc,Pastor:2002we,Balantekin:2004ug}, could potentially impact the
$r$-process. The case for this became much stronger in the last five years, since supercomputer calculations of
collective oscillations have become available (starting with \cite{Duan:2006an,Duan:2006jv}). These calculations have
shown that, even without sterile neutrinos, large flavor transformations can develop sufficiently close to the
neutrinosphere, where the $r$-process is thought to occur. Yet, with the exception of the recent attempt
\cite{Chakraborty:2009ej}, the impact of the new results on the $r$-process has not been investigated. 

The reason for this, we believe, is not the lack of interest, but the inherent complexity of the problem. First,
accurate modeling of collective oscillations requires supercomputers. This is so because -- unlike the more familiar MSW
flavor transformations, in which each neutrino evolves independently -- in the collective oscillation regime neutrinos
of different directions and different energies are coupled \cite{Qian:1994wh,Duan:2006an}. Second, one has a chain of
nuclear reactions to follow, as a function of the position in the outflow. Lastly, the physical conditions, in which the
candidate $r$-process takes place, themselves require detailed simulations.

In its full form, this is clearly a very ambitious program. The goal of the present paper is to gauge the effect and
establish whether or not its magnitude warrants further, more detailed modeling. Moreover, we would like to get a sense
of the size of the effect for a variety of conditions.  With these goals in mind, we pick astrophysical conditions which
produce neutron rich outflow, following models in the existing literature.  Since collective effects may cause rapid
oscillations above the neutrinosphere, we couple neutrino transformation calculations to a nuclear reaction network. We
present several examples of calculations obtained in this way. One may attempt to reduce the impact of the oscillations
on the $r$-process to a simple criterion, such at the electron fraction $Y_{e}$ at a certain radius. Such prescriptions
should also be applied with care and we discuss their limitations.

One is also tempted to simplify the treatment of the collective oscillations. We show, however, that such
simplifications may lead to entirely erroneous results.  We illustrate this with the example of the frequently used
``single-angle'' approximation, which in this case is seen to utterly fail.  This sensitivity to the details of the
oscillations -- and, correspondingly, the need to model this process carefully -- is one of the major results of this
paper. 

Lastly, although our outflow models are clearly simplified (and, indeed, as explained in the next Section, the exact
mechanism of the astrophysical $r$-process is yet to be conclusively established), we believe our basic framework and
findings should apply more generally. For example, NS-NS and BH-NS mergers also have neutrinos which, for hot outflows,
strongly influence the composition of the accretion disk and outflow \cite{Surman:2005kf,Met08}. We will return to this
important point at the end.

This paper is organized as follows. In Sect.~\ref{sect:overview}, we briefly summarize the current status of the
$r$-process modeling. In Sect.~\ref{sect:toy}, we illustrate the effects of neutrino flavor transformations using a toy
model. In Sect.~\ref{sec:collective}, we introduce the main physics of the collective transformations and give
order-of-magnitude arguments why these transformations may impact the $r$-process. In Sect.~\ref{sec:nuc_calc}, we
describe the nucleosynthesis network calculations and, in Sect.~\ref{sec:collectiverprocess} we present our sample
results. Finally, in Sect.~\ref{sec:conclusions}, we summarize our findings and discuss them in a more general context
of astrophysical nucleosynthesis.

\section{Overview of the $r$-process}
\label{sect:overview}

The abundance pattern of elements \cite{Bar05,Roe09} in the region between the $A=130$ peak and the
actinides is consistent with the rapid neutron capture process ($r$-process \cite{Bur57,Cam57}; see
\cite{Arn07} for a recent review). There is a long-standing question regarding the origin of the
$r$-process. Astrophysical data points to a phenomenon that happens early in the evolution of the
universe and has a relatively high frequency \cite{Sne96,Rya96,Ish99,Tru02}. Neutron star - neutron
star (NS-NS) mergers \cite{Mey89,Fre99} and black hole-neutron star (BH-NS) mergers
\cite{Surman:2008qf,Surman:2005kf} are difficult to reconcile with this data \cite{Arg04,WIs06}. On
the other hand, core collapse supernovae (CCSN) do have the right timescale.  This motivates a closer
look at possible $r$-process conditions in these objects.

Unfortunately, the exact mechanism by which CCSN produce an $r$-process has not yet been conclusively
established. The neutrino driven wind is a popular environment to study \cite{Mey92,Woo94}, but there
is by now a long series of papers in the literature that show that it is more difficult to produce the
requisite conditions than first thought, e.g.
\cite{Tak94,Ful95,McLaughlin:1997qi,Meyer:1998sn,Qia96,Hof97,Tho01,Ter02,Lie05}. 
 
Solutions to this problem may come from a better understanding of the relevant physics. To this end, as a necessary
step, one should identify and quantify all physical processes that may impact the $r$-process yields (whether or not
they solve the problem). Suggestions have been made, for instance, regarding neutrino physics
\cite{Wan06,Fry09,Fischer:2009af,Hue09,Gav09} and hydrodynamics \cite{Arc07,Far09}. As another example, fission cycling
could produce a good fit to the data \cite{Mar07,Beun:2007wf} but the electron neutrino fraction would have to be much
lower than is predicted by any model, e.g. \cite{Fischer:2009af,Hue09} for a recent example, save those that include
sterile neutrinos \cite{McLaughlin:1999pd,Fetter:2002xx,Cal00,Beun:2006ka}.

Our present investigation of the effects of neutrino oscillations may be viewed as part of this larger effort to
identify and quantify all relevant physics that could significantly impact the $r$-process yields. At the same time, we
wish to set up a framework, which should apply for a variety of conditions. Whatever the final solution to the
$r$-process is, one may need to address if and how the neutrino oscillations impact the yields.

\section{Toy model}
\label{sect:toy}

We begin our study of the influence of collective oscillations on the $r$-process by examining a toy
model of the supernova neutrino-driven wind. In the traditional picture of this environment, material
initially composed of free nucleons is heated to high entropy by the neutrinos emitted from the
cooling protoneutron star.  The heated material expands outwards and cools adiabatically; as the
temperature drops the nucleons assemble first into alpha particles and then into heavier nuclei.
Studies have shown that $r$-process nucleosynthesis can proceed in this environment if (a) there is a
neutron excess in the initial composition (i.e., the initial electron fraction $Y_{e}=1/(1+n/p)$ is
less than 0.5), (b) the entropy is sufficiently high to favor lighter nuclei at high temperatures, and
(c) the timescale is sufficiently fast to lower the efficiency of converting alpha particles to
heavier nuclei.  If the first condition is satisfied, the latter two conditions are necessary to
achieve the large neutron-to-seed ratio required for a robust $r$-process.  Note that this first
condition (a) is only a guideline and not a strict requirement, since it has been shown that in
certain cases $r$-process elements can be produced under proton rich conditions \cite{Meyer:2002zz}.

Neutrinos play an important role in each stage of the nucleosynthesis.  Weak interactions on nucleons
\begin{eqnarray}
\nu_{e} + n &\rightleftharpoons& p + e^{-}\\
\bar{\nu}_{e} + p &\rightleftharpoons& n + e^{+}
\end{eqnarray}
are responsible for setting the electron fraction $Y_{e}$ in the ejected material.  
In regions most relevant for nucleosynthesis, electron neutrino captures on neutrons and electron
antineutrino captures on protons dominate. If the electron antineutrino spectrum is hotter than the
electron neutrino spectrum, as is traditionally assumed for the late-time neutrino emission from a
cooling protoneutron star, then the rate of electron antineutrino captures will be faster and a
composition with more neutrons than protons will be favored.  The neutrino physics is therefore
responsible for the initial neutron excess crucial for the $r$-process.   

When the temperature in the outflow drops to the point where nuclear reassembly begins, all of the
protons (and a corresponding number of neutrons) become locked up in alphas, shutting off the electron
antineutrino captures on protons that produce free neutrons.  At this point, electron neutrino
interactions on neutrons act to reduce the available free neutron abundance directly and also
indirectly, as the protons produced by the neutrino captures bind with additional free neutrons to
form alpha particles.  This is the so-called `alpha effect' detailed in
\cite{Ful95,McLaughlin:1997qi,Meyer:1998sn}.  After the epoch of alpha particle formation is past,
electron neutrino interactions continue to reduce the availability of free neutrons but the effect
weakens as the material moves away from the protoneutron star.

If the traditional hierarchy of neutrino energies is assumed, i.e. $\langle E_{\nu_{\mu,\tau}}\rangle >
\langle E_{\bar{\nu}_{e}}\rangle > \langle E_{\nu_{e}}\rangle$, then $e, \mu, \tau$ flavor
oscillations \emph{may} act to raise the effective energies of the $\nu_{e}$ and $\bar{\nu}_{e}$
fluxes.  This has important nucleosynthetic consequences, with the severity determined by which the
stage the nucleosynthesis is in when the flavor transformation occurs.  If the transformation happens
at early times, when the composition of the material is mostly free nucleons, the electron fraction
will readjust.  A transformation that takes place when the composition is dominated by alphas will
enhance the alpha effect, and a transformation that occurs after seeds form will reduce the number of
free neutrons available for capture during the $r$-process. 

Of course, the impact of the oscillations should depend on the details of the process, particularly on which parts of
the spectra are permuted and at which distances. This requires detailed modeling of the oscillations, which is presented
later. Here, we would like to illustrate these effects, assuming the oscillations lead to a simple swap of entire
spectra between flavors. 

We present the results from such a toy calculation in Fig.~\ref{fig:toy}.  We
took the neutrino fluxes to be the late-time fluxes from \cite{Kei03}, specifically the simulation
labeled $q=3.0$. It describes spectra of the form $\propto E^{2}[1+\exp(E/T-\eta)]^{-1}$, with
neutrino temperatures $T_{\nu_{e}}=2.6$, $T_{\bar{\nu}_{e}}=4.0 \, {\rm MeV}$, and $T_{\nu_{\mu}}=5.0
\, {\rm MeV}$, degeneracy parameters $\eta_{\nu_{e}}=3.0$, $\eta_{\bar{\nu}_{e}}=2.8$, and
$\eta_{\nu_{\mu}}=1.8$, and luminosities $L_{\nu_{e}}=6.6\times 10^{51}$ ergs s$^{-1}$,
$L_{\bar{\nu}_{e}}=8.8\times 10^{51}$ ergs s$^{-1}$, and $L_{\nu_{\mu}}=12.7\times 10^{51}$ ergs
s$^{-1}$.  These neutrino parameters set the initial electron fraction of the material at
$Y_{e}=0.41$. (Note that recent calculations of proto-neutron star evolution predict neutrino
parameters that produce higher electron fractions and would require more extreme conditions to produce
$r$-process elements than those used here \cite{Fischer:2009af,Hue09}.)  We chose a sample hydrodynamic
trajectory that results in a main $r$-process for the given neutrino parameters.  We then ran four
simulations, one with neutrino interactions turned off after the electron fraction was set, one with
unchanged neutrino fluxes throughout, and the remaining two with toy neutrino swaps, where the
electron neutrino flux was simply replaced by the $\mu$ neutrino flux at a specified radius.  A
comparison between the simulations with unchanged neutrino physics and with no neutrino interactions
for $T<10^{10}$ K show the operation of the normal alpha effect, whereby neutrino interactions on
neutrons result in more alphas and fewer neutrons (top panel of Fig.~\ref{fig:toy}) and a less robust
$A\sim 195$ peak in the final $r$-process abundance distribution (bottom panel of Fig.~\ref{fig:toy}). 
A neutrino flavor swap that takes place after alpha particle formation acts to enhance the alpha
effect; the toy swap increases the alpha particle mass fraction, decreases the neutron abundance, and
weakens the resulting $r$-process.  A neutrino flavor swap that takes place while the composition is
dominated by free nucleons causes a readjustment of the neutron-to-proton ratio and an even stronger
alpha effect, with a dramatic influence on the subsequent $r$-process, which in this case does not go
beyond the $A\sim 130$ peak.

\FIGURE[p]{\epsfig{file=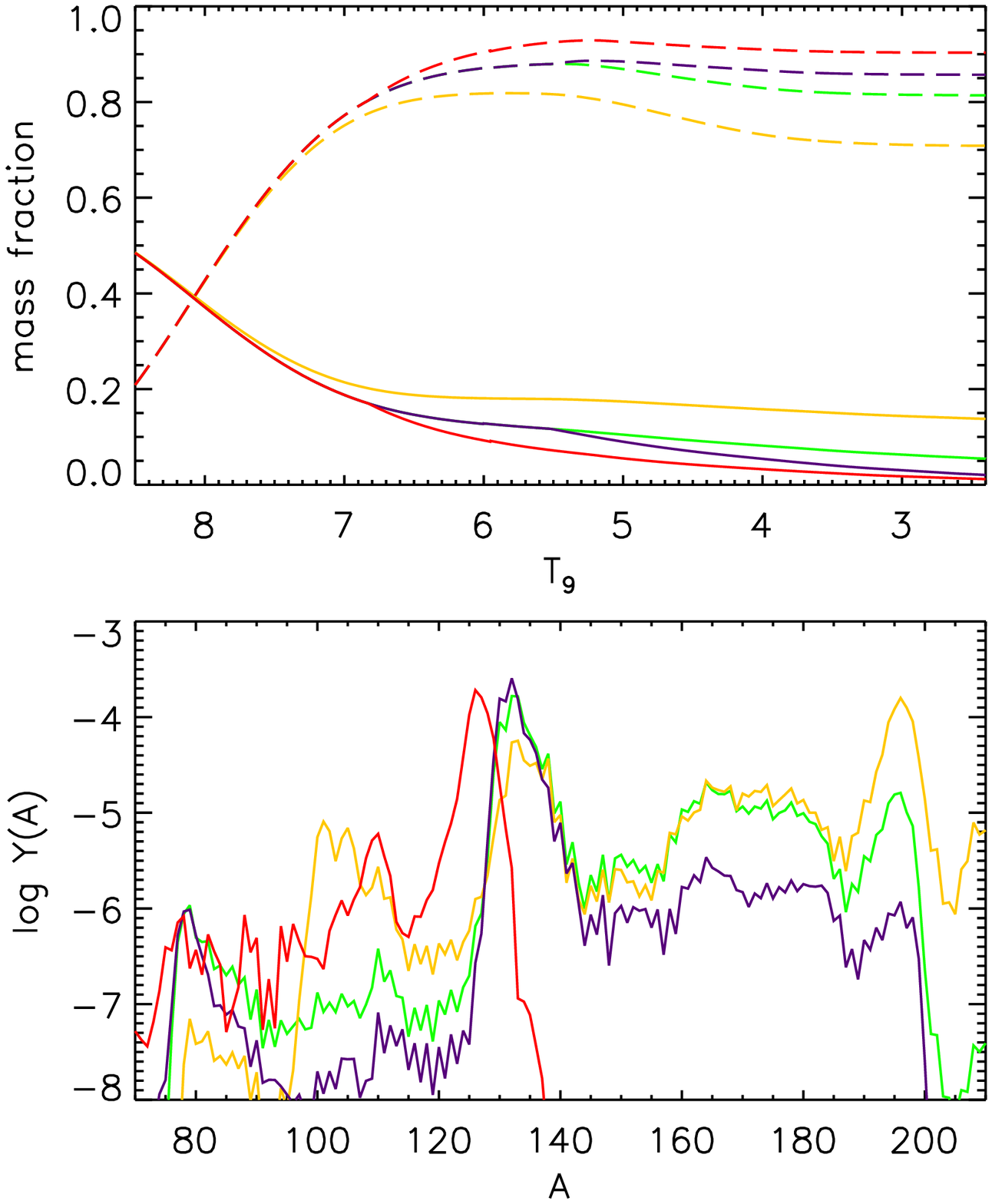,width=14cm}
        \caption{The upper panel shows the mass fractions of neutrons (solid lines) and alpha particles
(dashed lines) for sample simulations with no neutrino oscillations (green), a test neutrino swap that
occurs at the point of alpha particle assembly into heavier nuclei (purple), a test neutrino swap that
occurs during assembly of nucleons into alphas (red), and neutrino interactions turned off for $T_{9}<9$
K (yellow).  The bottom panel shows the final $r$-process abundances for each of the simulations.  All
simulations use the same hydrodynamic trajectory, with a timescale of 20 ms and an entropy of
$s/k=200$, and the neutrino fluxes are the $q=3.0$ case from \cite{Kei03}.}
        \label{fig:toy}}

Our toy model confirms that the influence of the neutrino flavor transformations on supernova nucleosynthesis depends
sensitively on what stage of nucleosynthesis the material is in when the transformation occurs.  This is determined by
the hydrodynamic trajectory of the outflowing material.  Fig.~\ref{fig:pe} shows approximate temperature ranges for each
stage of the nucleosynthesis along with the temperature as a function of radius for two sample hydrodynamic
trajectories, described in Section \ref{sec:nuc_calc}.  Fig.~\ref{fig:pe} also shows a sample electron neutrino survival
probability $P(e)$ for 20 MeV neutrinos, calculated as described in Section \ref{sec:collective} below.  In this
example, the maximum rate of the flavor transformation occurs at roughly 140 km from the neutron star.  For the lower
entropy trajectory shown by the dashed line, this corresponds to the assembly of seed nuclei, and so we anticipate the
nucleosynthetic consequences to be similar to that shown by the purple line in the toy model plot, Fig.~\ref{fig:toy}. 
For the higher entropy trajectory given by the dot-dashed line in Fig.~\ref{fig:pe}, the same location corresponds to an
earlier stage of nucleosynthesis, namely, the assembly of alpha particles.  The influence of the flavor transformation
is expected to be more dramatic here, closer to that shown by the red line of Fig.~\ref{fig:toy}.  Since large
uncertainties remain in the astrophysical conditions of the $r$-process, we investigate the nucleosynthesis using a
sampling of parameterized wind trajectories in which the swap may occur during earlier or later stages in the
nucleosynthesis.

\FIGURE[p]{\epsfig{file=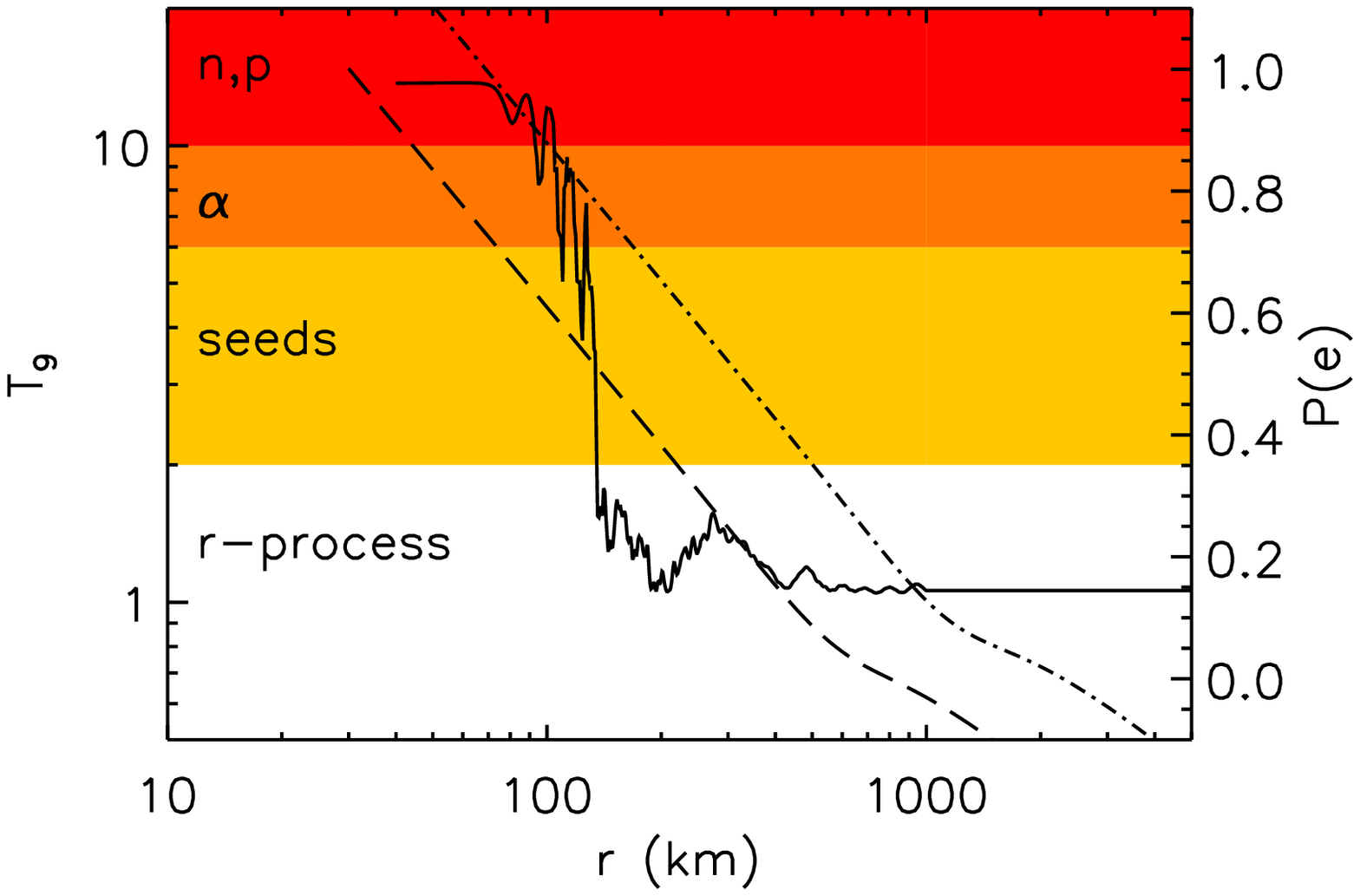,width=14cm}
           \caption{Shows temperature $T_{9}$ as a function of radius in km (dashed line) for two sample thermodynamic
trajectories, one with lower density and $s/k=300$ (dot-dashed line) and the
other with higher density and $s/k=200$ (dashed line); the horizontal bands show the approximate
temperature ranges for each stage of the nucleosynthesis as labeled. The solid line shows a sample electron neutrino
survival probability $P(e)$ for 20 MeV neutrinos, as calculated assuming an inverted hierarchy.  For the trajectory with
the early-type density profile, the flavor transformations occur during alpha particle formation, while for the
trajectory with the late-type density profile the flavor transformations occur later, during the $r$-process seed
assembly.}
           \label{fig:pe}}

\section{Collective neutrino oscillations}
\label{sec:collective}

Let us now turn to modeling neutrino flavor oscillations. As we mentioned earlier, the main mechanism of flavor
conversion that is relevant to $r$-process nucleosynthesis is the collective oscillations, caused by coherent
neutrino-neutrino interactions, rather than the more familiar MSW (matter-enhanced) conversion. We begin by reviewing
the two processes.

Outside of the neutrinosphere, the neutrinos by definition do not experience any incoherent scattering. Yet, their
\emph{coherent forward} scattering on the background particles remains important for the flavor evolution. The MSW
effect \cite{Wolfenstein:1977ue,Mikheev:1986gs} is the textbook example of this phenomenon. The scattering involved in
this process is on electrons, protons, and neutrons of the medium. The MSW process is well known to operate in the Sun
and should also operate in the supernova environment, as was recognized early on \cite{Mikheev:1986if}. Yet, for the
conditions typical of an iron core supernova, this process does not impact the $r$-process. 

To see this, consider the condition for the transformation, which is given by the comparison of the vacuum and matter
terms in the neutrino oscillation Hamiltonian,
\begin{eqnarray}
\label{eq:msw}
\Delta m^{2}/2E_{\nu} \sim \sqrt{2} G_{F} N_{e}.
\end{eqnarray}
The l.h.s. is either of the two eigenvalue splittings in the vacuum oscillation term; the r.h.s. is the corresponding
splitting in the Wolfenstein matter term.  Here, $E_{\nu}$ is the neutrino energy, $N_{e}$ is the number density of
background electrons and $\Delta m^{2}$ is either $\sim 7.7\times 10^{-5}$ eV$^{2}$, or $\sim 2.7\times 10^{-3}$
eV$^{2}$, corresponding to the ``solar'' and ``atmospheric'' mass splittings. Both $\Delta m^{2}$ values lead to flavor
transformations, although the details of the transformations differ
\footnote{In the limit of the small mixing angle, the MSW transformation occurs either in neutrinos or in antineutrinos (depending on where the resonance is, which, in turn, depends on the sign of the mass hierarchy). This regime applies to the ``atmospheric'' transformation, since it is controlled by the small $\theta_{13}$ angle. In contrast, the ``solar'' mixing angle is large and \emph{both} neutrinos and antineutrinos transform, at similar densities. The notion of ``resonant'' vs. ``nonresonant'' transformation is not physically useful here. To this end, note that Eq.~(\ref{eq:msw}) does not contain the factor of $\cos2\theta$ that is present in the traditional definition of the resonance \cite{Friedland:2000rn}.}. 

The important point is that in an iron core collapse supernova, either ``atmospheric'' or ``solar'' MSW transformations
typically occur at large radii, beyond those of interest for the $r$-process. Indeed, it follows from Eq.~(\ref{eq:msw})
that for typical neutrino energies of $\sim10$ MeV and the electron fraction of $Y_{e}\sim0.4$, the ``atmospheric''
transformation takes place at densities of $\sim10^{3}$ g/cm$^{-3}$ and the ``solar'' one at still lower densities.
Comparing this with the wind profiles in Fig.~\ref{fig:profile}, we see that the MSW transformations occur at $r\gtrsim
500$ km, by which point even the formation of the seeds is complete (Fig.~\ref{fig:pe}).

Let us now turn to collective oscillations. Let us note at the outset that the physics of this phenomenon has been shown
to be very rich. The subject is still being actively explored and even a complete discussion of the established results
would be well beyond the scope of the present paper. The reader is encouraged to consult specialized literature on the
subject ({\it e.g.}, \cite{Duan:2009cd,Duan:2010bg} and references therein). Our goal here is to introduce the basic
relevant physics and explain, using simple order-of-magnitude considerations, why this effect could be relevant to the
$r$-process.

The starting observation is that during the first ${\cal{O}} (10)$ seconds of the explosion the number density of
streaming neutrinos $N_{\nu}$ in a supernova is very high. Because of this, close to the neutrinosphere, their mutual
coherent scattering is physically important
\cite{Fuller:1987,Notzold:1987ik,Pantaleone:1992xh,Pantaleone:1992eq,Sigl:1992fn,McKellar:1992ja}. In addition to the
vacuum oscillation term and the Wolfenstein matter term, the oscillation Hamiltonian in this regime contains the term
induced by the neutrino background \cite{Pantaleone:1992xh,Pantaleone:1992eq},
\begin{equation}\label{eq:Pantaleone_term}
    H_{\nu\nu}^{(j)} = \sum_{i} \sqrt{2} G_{F} N_{\nu} (1-\cos\Theta_{ij})|\psi_{i}\rangle\langle\psi_{i}| - \sum_{\bar i} \sqrt{2} G_{F} N_{\bar\nu} (1-\cos\Theta_{\bar i j})|\psi_{\bar i}^{\ast}\rangle\langle\psi_{\bar i}^{\ast}|.
\end{equation}
The index $i$ runs over all neutrino groups and $\bar i$ over all antineutrino groups. The index $j$ refers to the
neutrino group for which the equation is written: different groups are characterized by different energies and/or
directions. The interaction strength between any two groups depends of the angle between particle momenta, $\Theta_{ij}$
($\Theta_{\bar ij}$).

The evolution Hamiltonian for each neutrino group now contains contributions from all neutrinos in the ensemble.
Therefore, flavor evolutions of neutrinos and antineutrinos, of different flavors, energies and on different
trajectories become coupled. Thus, despite being superficially similar to the Wolfenstein matter term, the
neutrino-neutrino term, where it is important, brings about qualitative change to the problem. The neutrino ensemble now
evolves \emph{collectively}. 

What is the condition for the neutrino-neutrino term to be physically important? By analogy with Eq.~(\ref{eq:msw}), one
may start by comparing the dimensionful coefficient in Eq.~(\ref{eq:Pantaleone_term}) to those of the neutrino-matter
interactions and the vacuum oscillation term. This gives ({\it e.g.}, \cite{Friedland:2003eh}),
$|N_\nu-N_{\bar\nu}|\gtrsim |N_{e^-}-N_{e^+}|$ and $G_F|N_\nu-N_{\bar\nu}|\gtrsim \Delta m^2/E_\nu$, correspondingly.

The next step is to account for the angular factors in Eq.~(\ref{eq:msw}), which are important for the conditions
specific to the supernova. Indeed, at sufficiently large distances from the neutrinosphere neutrino rays becomes nearly
collinear and therefore the geometric factors $(1-\cos\Theta)$ become parametrically small. Comparing the neutrino
self-interaction term to the vacuum oscillation term, on gets
\begin{eqnarray}
\label{eq:nunu}
G_F |N_\nu-N_{\bar\nu}|\langle1-\cos\Theta(r_{\nu\nu})\rangle\gtrsim\Delta m^2/E_\nu.
\end{eqnarray}
Here, the brackets denote an appropriate averaging over angles.  The l.h.s. falls rapidly with radius, as $\propto
N_{\nu}(r) (1-\cos\Theta) \propto r^{-4}$ \cite{Pantaleone:1994ns}, suggesting that any transformations have to occur
close to the neutrinosphere. Indeed, given the late-time emission rate of $\sim10^{51}\mbox{ erg/s}/10\mbox{ MeV}\sim
10^{56}$ neutrinos per second, the radius of the neutrinosphere $R_{\nu}\simeq10$ km, and the atmospheric $\Delta
m^{2}=2.7\times10^{-3}$ eV$^{2}$, Eq.~(\ref{eq:nunu}) gives $r_{\nu\nu}\sim{\cal O}$(100 km). 

Importantly, a similar geometric correction does not apply when comparing the neutrino self-interaction term to the
background matter term (which would suppress collective oscillations). This is because the overall effect of the
background matter can be ``rotated away'', as explained in \cite{Duan:2005cp}. This important physical result is the
reason why large collective flavor transformations do occur under realistic supernova conditions
\cite{Duan:2006an,Duan:2006jv}. The residual dispersion of the matter phase on different trajectories, however, cannot
be rotated away, so that very dense matter does suppress collective oscillations \cite{EstebanPretel:2008ni}. The
comparison of this dispersion to neutrino self-interaction strength shows the inequality\begin{eqnarray}
\label{eq:nunumatter}
|N_\nu-N_{\bar\nu}|\gtrsim|N_{e^-}-N_{e^+}|
\end{eqnarray}
is unmodified \cite{EstebanPretel:2008ni}.

Note that the same argument applies to the vacuum oscillation term, but in this case the dispersion with energy is of
the same size as $\Delta m^{2}/E_{\nu}$. Still, strictly speaking, it should be understood that the r.h.s. of
Eq.~(\ref{eq:nunu}) contains \emph{the dispersion} of the vacuum term (with antineutrinos counting as negative
energies).
 
Lastly, one should also consider the dispersion of the neutrino self-interaction on different trajectories. It can be
easily shown that, similarly to the matter term, this dispersion is comparable to the average strength of the
interaction \cite{Duan:2010bf}. In the frequently used \emph{single-angle} approximation, this dispersion is neglected
by fiat (and the calculation considerably simplifies as a result). The justification for this approximation came not
from any first principles argument, but rather from comparing the results of single-angle and multiangle calculations in
certain setups. On the other hand, as shown in \cite{Duan:2010bf}, single-angle calculations do not always reproduce
full multiangle results. In cases when they disagree, the dispersion in neutrino self-interaction term can lead to the
suppression of the oscillations close to the neutrinosphere. The condition for the oscillations is
\begin{eqnarray}
\label{eq:nunu2}
|N_\nu-N_{\bar\nu}|\langle1-\cos\Theta(r_{\nu\nu})\rangle\sim\Delta m^2/(G_F E_\nu),
\end{eqnarray}
{\it e.g.}, the inequality in Eq.~(\ref{eq:nunu}) becomes an approximate equality; flavor transformations occur just
before the neutrino-neutrino interaction ceases being physically important ({\it cf} Eq.~(\ref{eq:nunu})). This
multiangle suppression proves crucial for the analysis of the $r$-process, as we show in what follows. 

To recapitulate, Eqs.~(\ref{eq:nunu}) and (\ref{eq:nunumatter}) give the \emph{necessary} conditions for the neutrino
flavor transformations, if any, to be in the collective regime. This equations, however, do not tell what that evolution
is: this is the question of dynamics.  With this in mind, we finally turn to our modeling of the oscillations. 

Since we are interested in the $r$-process in the neutrino driven wind environment, we specialize to the late-time
spectra, taken from \cite{Kei03}. As our baseline scenario, we choose the $q=3.0$ point from Table 6 in that paper, as
described in the previous Section. The matter density profiles we use are shown in Fig.~\ref{fig:profile}. The densities
are low enough so that the matter suppression mechanism does not delay the onset of the oscillations. On the other hand,
the anisotropy-driven (multiangle) suppression is present for these spectra. While in the single-angle calculations the
oscillations start right away, as the neutrinos are released, in the multiangle treatment the oscillations are delayed.

This is qualitatively similar to what is described in \cite{Duan:2010bf}, although it must be stressed that the
quantitative details are different, in ways that are important for the $r$-process. Unlike in \cite{Duan:2010bf}, where
the multiangle calculations predicted oscillations starting at $\sim125$ km, here we find the oscillations start at
$\sim75$ km. This difference is due to the choice of the initial spectra: here, as mentioned above, we use the $q=3.0$
spectra from Table 6 in \cite{Kei03}, while the computations in \cite{Duan:2010bf} are done for the $q=3.5$ spectra. We
consider the $q=3.0$ spectra here because they create conditions more suitable for the $r$-process.

We consider oscillations between all three known neutrino flavors, since three-flavor evolution may be entirely
different from the two-flavor approximation ({\it e.g.}, \cite{Friedland:2010sc}). We use the known solar and
atmospheric neutrino oscillation parameters, set the value of the unknown angle $\theta_{13}$ to 0.01, and consider both
signs of the neutrino mass hierarchy. The starting radii for the oscillations with our baseline spectra happen to be the
same for the normal and inverted mass hierarchies. The oscillation patterns, however, develop differently in the two
cases and therefore the impact on the nucleosynthesis needs to be considered separately. The dependence on the hierarchy
becomes even more obvious as one varies the spectra. As an illustration, we vary the luminosities of the non-electron
neutrino flavors, $L_{x}$. With $L_{x}$ suppressed by a factor of 0.7, for example, the starting radius for oscillations
becomes 70 km for the inverted hierarchy, while for normal hierarchy the collective oscillations disappear altogether.
This sensitivity of the starting radius on the details of the assumed spectra is, in fact, an important generic feature
of the collective oscillations in this regime. We will show how these changes impact the $r$-process predictions.

Lastly, a word of caution: one should exercise care when applying oscillation results available in the literature to the
$r$-process calculation. Most papers on collective oscillations are typically focused on modeling the neutrino signal in
a terrestrial detector.  Correspondingly, the neutrino spectra are typically reported at sufficiently large radii,
chosen so that the transformations have completed. Typically, at this stage one finds a pattern of spectral ``swaps'',
{\it i.e.}, neutrinos of different flavors exchange spectra in certain energy intervals.  It would be completely
incorrect to use these final spectra to model the $r$-process. Indeed, the swaps are formed only at hundreds of
kilometers from the collapsed core. What we need instead is the detailed information of how the oscillations develop
early on, not what they finally settle to. That is why we compute our oscillations ``from scratch''.

\section{Nucleosynthesis calculation}
\label{sec:nuc_calc}

For the full nucleosynthesis calculations, we begin with a nuclear statistical equilibrium calculation
\cite{McLaughlin:1999pd} to follow the composition of free nucleons and alphas at high temperature,
switch to an intermediate mass nuclear network calculation \cite{Hix99} to determine the mass
fractions of the seed nuclei, and finally finish with a simplified nuclear network calculation
(including neutron capture, photodissociation, beta decay, beta-delayed neutron emission, and
charged-current neutrino interactions) \cite{Sur01} to find the final $r$-process abundances. 

For our studies we use a modified neutrino driven wind model. Neutrino driven wind models were first
developed by \cite{Duncan:1986} and were then improved upon by
\cite{Qian:1996xt,Cardall:1997bi,Thompson:2001ys,Terasawa:2004ju}.  Recently, it has been shown that
the traditional exponential outflow should be modified by a slowdown as the neutrino heated material
reaches the supernova shockfront \cite{Arc07}.

Here we use a range of adiabatic wind trajectories based on the parameterization of Panov
and Janka \cite{Pan09}.  The parameterization in \cite{Pan09} takes the first stage of the outflow to
be a steady-state expansion with an exponentially-growing radius, $r(t) = r_{\mathrm{init}}
e^{t/\tau}$, and an exponential decline in density, $\rho = \rho_{\mathrm{init}} e^{-t/ 3 \tau}$,
where $\tau$ is the initial dynamical timescale.  In the second stage of the outflow, corresponding to
its evolution after deceleration by a reverse shock, the density declines more slowly, as a power law
$\rho=\rho_{0} (t/t_{0})^{-2}$, where $\rho_{0}$ and $t_{0}$ are the density and time at the point of
transition to the power law dependence, taken in \cite{Pan09} to be the time at which the temperature
is $T_{9}=1$.  The steady-state condition $r^{2}\rho v=\mathrm{constant}$ is used to find $r(t)$ and
$v(t)$.

We use this parameterization to calculate two density profiles $\rho(r)$, shown in
Fig.~\ref{fig:profile}, that are roughly consistent with, but not fit to, early-type and late-type
density profiles from Arcones {\it et al} \cite{Arc07}.  We then generate a range of hydrodynamic and
thermodynamic trajectories from these profiles by varying entropy and initial timescale.  The constant
entropy is used to calculate temperature from the density, and the initial timescale determines the
evolution of the velocity, which is calculated from the steady-state condition $r^{2}\rho
v=\mathrm{constant}$.  Note that this parameterization differs from that of \cite{Pan09} only in how
the switchover point from exponential to power law is determined; here it is not set by a temperature
condition but from the density profile.  We examine trajectories with a variety of entropies,
$s/k=100$, 200, 300, and a range of initial timescales, 1 ms $< \tau <$ 50 ms.

\FIGURE[p]{\epsfig{file=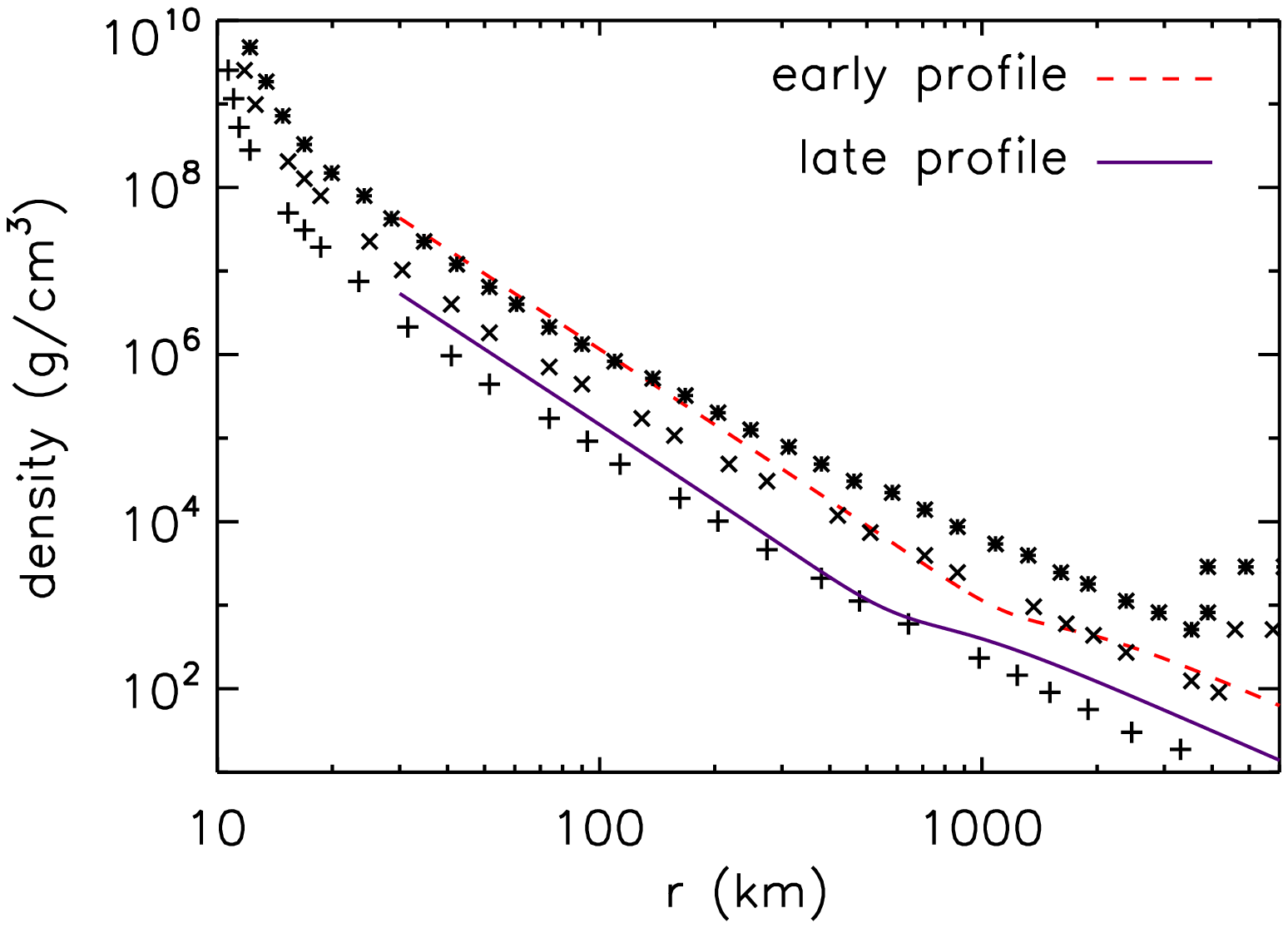,width=14cm}
          \caption{Shows two sample density profiles---an early-type profile (dashed line) and a late-type profile
(solid line)---used in this work. To show the similarities between our parameterized profiles and recent
hydrodynamical calculations, we also include profiles extracted from Arcones {\it et al} \cite{Arc07}, corresponding to
$t=2$ s (asterisks), $t=4$ s (x's), and $t=10$ s (crosses) after core bounce.}
          \label{fig:profile}}

The neutrino fluxes and survival probabilities calculated as described above were output every kilometer
from 40 to 1000 km for the nucleosynthesis calculation.  The electron neutrino and antineutrino capture
rates per nucleon that result are shown in Fig.~\ref{fig:rates} as a function of radius, compared to
the rate assuming no oscillations.  While the neutrino capture rates are unchanged close to the neutron
star, collective oscillations increase the rates farther out, by up to factors of two to three.  

\FIGURE[p]{\epsfig{file=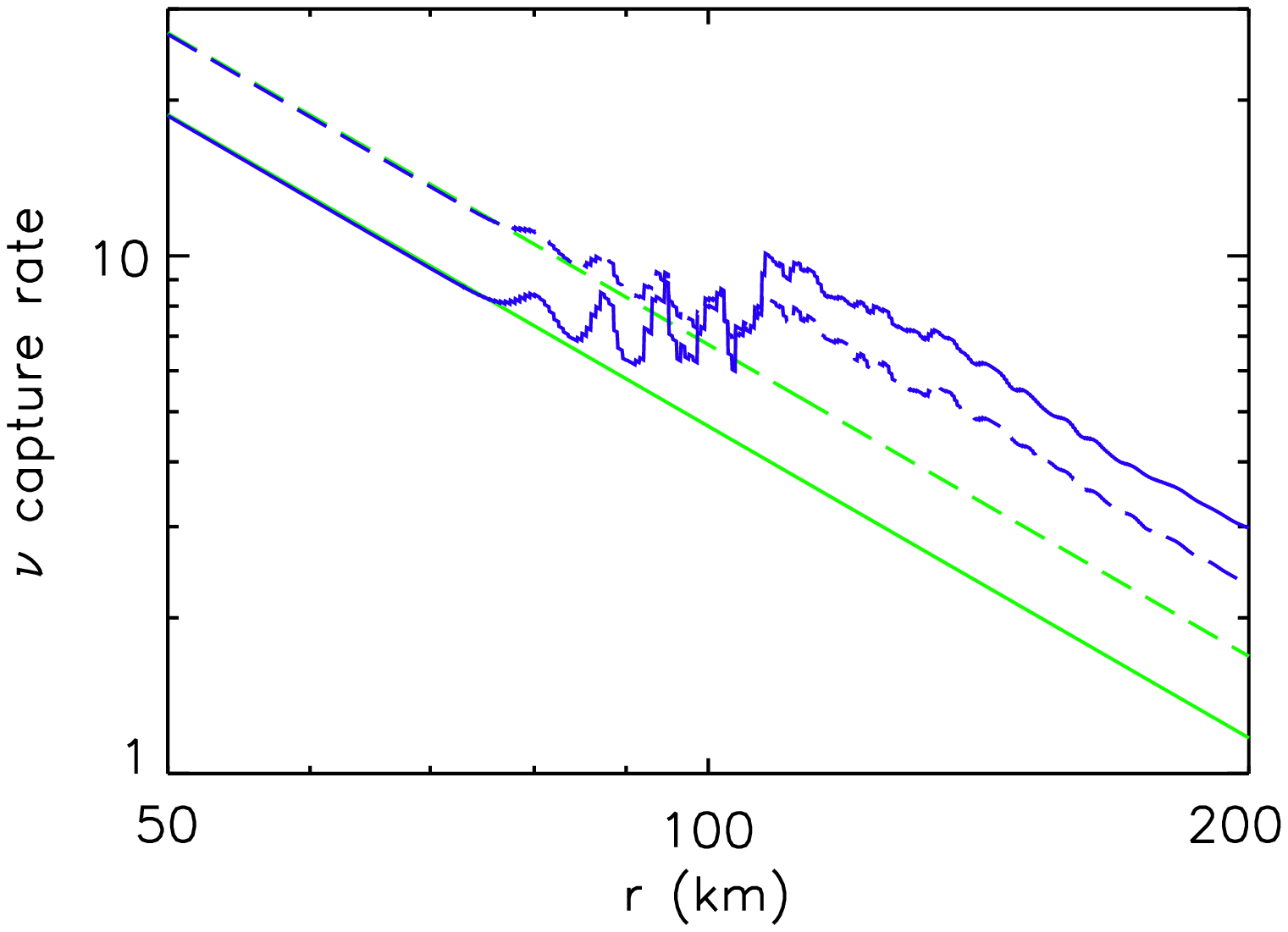,width=14cm}
          \caption{Shows electron neutrino (solid line) and electron antineutrino (dashed line) capture
rates per nucleon as a function of radius in km for the neutrino fluxes with no oscillations (green)
and with the full multiangle oscillation calculation (blue), assuming an inverted hierarchy for the
neutrino masses.  The neutrino fluxes are the $q=3.0$ case from \cite{Kei03}, with neutrino
temperatures $T_{\nu_{e}}=2.6$, $T_{\bar{\nu}_{e}}=4.0$, and $T_{\nu_{\mu}}=5.0$, degeneracy parameters
$\eta_{\nu_{e}}=3.0$, $\eta_{\bar{\nu}_{e}}=2.8$, and $\eta_{\nu_{\mu}}=1.8$, and luminosities
$L_{\nu_{e}}=6.6\times 10^{51}$ ergs s$^{-1}$, $L_{\bar{\nu}_{e}}=8.8\times 10^{51}$ ergs s$^{-1}$, and
$L_{\nu_{\mu}}=12.7\times 10^{51}$ ergs s$^{-1}$.}
          \label{fig:rates}}

We note that previous work \cite{Balantekin:2004ug,Chakraborty:2009ej} on the influence of collective
oscillations on nucleosynthesis have relied upon calculations of the equilibrium electron fraction to provide
a first estimate of the type of nucleosynthesis likely in the outflow.  This equilibrium electron fraction is
approximated as $Y_e = 1/\left(1 + {\lambda_{ \bar{\nu}_e} \over \lambda_{\nu_e}}\right)$. This equation comes
from the more general \cite{McLaughlin:1997qi}
\begin{eqnarray}
\label{eq:yedot}
\dot{Y}_e  &=&   (\lambda_{\nu_e,n} + \lambda_{e^+,n}) Y_n - (\lambda_{\bar{\nu}_e,p} + \lambda_{e^-,p}) Y_p 
   \\ 
 & & \mbox{}
 +   \sum_{A(N,Z)} (\lambda_{\nu_e, A(N,Z)} + \lambda_{e^+,A(Z,N)} - \lambda_{\nu_e, A(N,Z)} - 
    \lambda_{\bar{e}^+,A(Z,N)}) Y_{A(N,Z)} \nonumber 
\end{eqnarray}
In general this must be solved together with the other nuclear reactions that occur during nucleosynthesis. To
find the approximate form of $Y_e$ from Eq. \ref{eq:yedot}, one assumes that the rate of change of conditions
is relative slow compared to the weak rates so that $\dot{Y}_e \approx 0$.  It assumes also that electron
capture can be neglected and that there are no nuclei at all in the composition.  For $r$-process scenarios
this sort of approximation is useful when estimating the size of effects that operate before the composition
shifts from nucleons to nuclei, such as the effect of weak magnetism \cite{Horowitz:2001xf}.  This, however,
is not the case for flavor transformation.  Fig.~\ref{fig:ye} shows a comparison between the equilibrium
electron fractions to the electron fractions as determined in the full nucleosynthesis calculation for cases
both with and without oscillations.  The equilibrium calculations of $Y_{e}$ deviate from the full
calculations even early on.  By the time the swap occurs, the weak reactions have slowed and the composition
has shifted from nucleons to nuclei, and so the equilibrium $Y_{e}$ no longer provides even an approximate
estimate of the nucleosynthetic outcome.

\FIGURE[p]{\epsfig{file=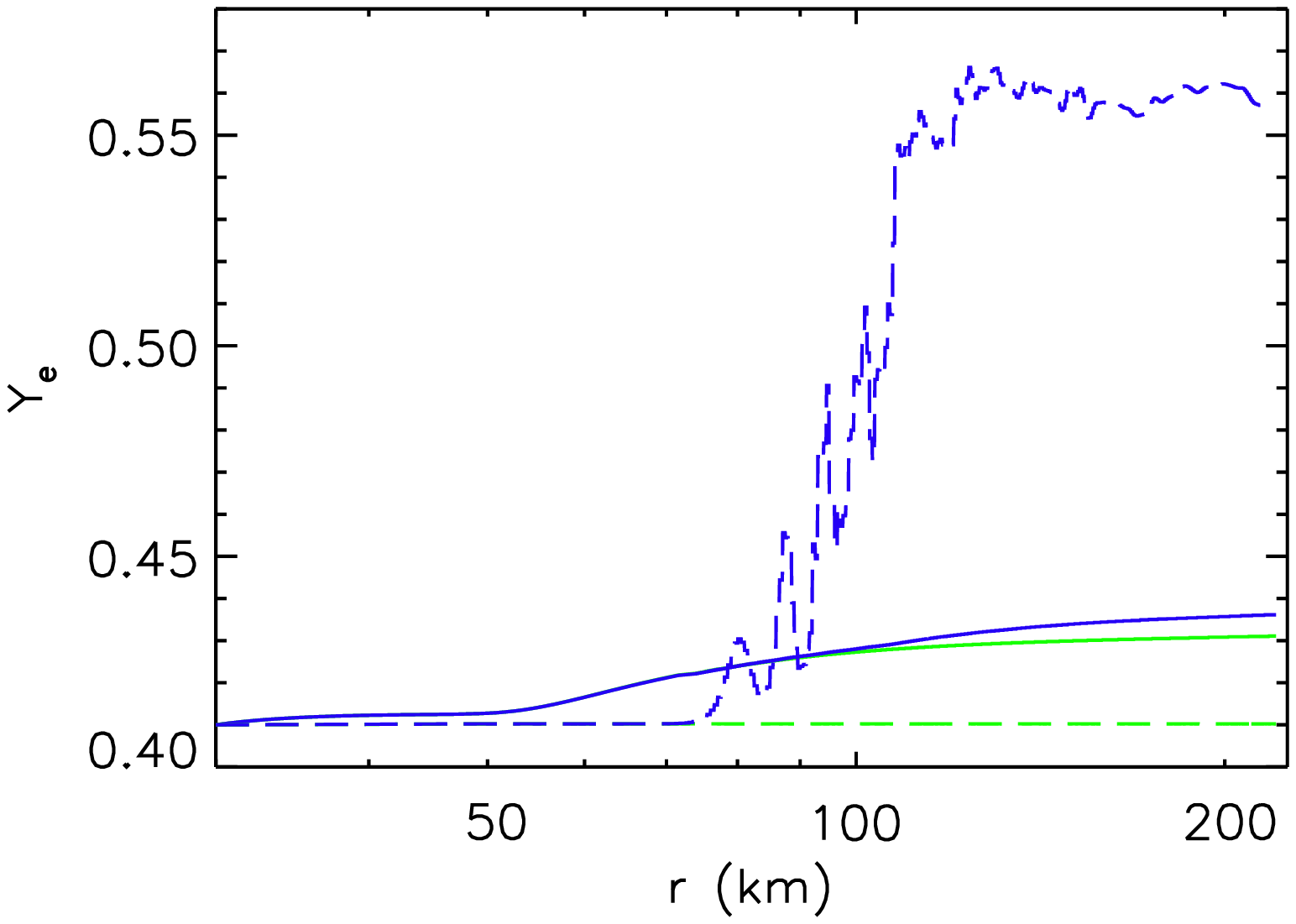,width=14cm}
          \caption{Shows the electron fraction as a function of radius in km for the full nucleosynthesis
calculation described in Sec. 4 for an example hydrodynamic trajectory and varying treatments of the neutrino
physics---no oscillations (green line) and oscillations in the full multiangle calculation (blue line).  These
are compared to the electron fractions calculated using the approximate formula $Y_e = 1/\left(1 + {\lambda_{
\bar{\nu}_e} \over \lambda_{\nu_e}}\right)$ for cases with no oscillations (dashed green line) and with the full
multiangle oscillation calculation (dashed blue line).  All oscillation calculations here assume an inverted
hierarchy for the neutrino masses, and the neutrino rates used here are those of Fig.~\ref{fig:rates}.}
          \label{fig:ye}}

\section{Collective oscillations in a full nucleosynthesis network calculation}
\label{sec:collectiverprocess}

Our simulations using the full neutrino oscillation calculation results with the nuclear network calculation as
described above support the qualitative conclusions from the toy model.  As described above, we ran a set of simulations with hydrodynamic trajectories using two density profiles, each with a variety of entropies, $s/k=100$, 200, 300, and a range of initial timescales, 1 ms $< \tau <$ 50 ms.  In this section we describe a
few cases that illustrate the overall effects of collective oscillations observed in this set of simulations.

We begin by examining the consequences of collective neutrino flavor transformation in our model, and comparing
the results of single angle to the results of multiangle calculations.  Fig.~\ref{fig:YvAa} shows sample final
$r$-process abundance distributions versus mass number $A$ from simulations using the early-time density profile
with entropy $s/k=200$ and timescale $\tau=15$ ms.  The simulations have the same hydrodynamic trajectory but
varying neutrino physics: one simulation assumes no neutrino oscillations, a second simulation incorporates
oscillations calculated in the single-angle approximation, and the third uses a full multiangle calculation for
the oscillations.  These are compared to a simulation where neutrino interactions are turned off at $T_{9}\sim
9$.  Note that the influence of the flavor transformations here is to enhance the influence of neutrino
interactions on the $r$-process and further weaken the third $r$-process peak.  In the single-angle calculation,
the third peak is entirely absent. 

These results can be understood by examining the early-stage nucleosynthesis depicted in Fig.~\ref{fig:nuc},
which shows the evolution of the free neutron abundance and the alpha particle mass fraction through the
formation of $r$-process seeds.  In the multiangle calculation, the flavor transformation occurs around $r\sim
100$ km or so, late in the alpha particle formation stage, so here the oscillations tend to intensify an alpha
effect that also exists in the case without oscillations.  This multiangle case is most similar to the purple
line in the toy model shown in Fig. \ref{fig:toy} , where a swap was introduced at the time when heavy nuclei
begin to form. When using the oscillation calculation with the single angle approximation, however, the flavor
transformation occurs much earlier, while the initial neutron-to-proton ratio is still being set.  This results
in a sufficiently large drop in the free neutron abundance to prevent the formation of the third $r$-process
peak.  This latter case is similar to the red line in Fig. \ref{fig:toy}, where a swap was introduced at the
beginning of alpha particle formation. In this situation, the inclusion of collective oscillations makes an
(erroneously) large impact on the abundance distribution.  This single angle approximation is inadequate because
it predicts that the oscillation begins too early in the nucleosynthetic process: the composition of the
material at the point where the oscillation begins plays a major role in determining the outcome.

\FIGURE[p]{\epsfig{file=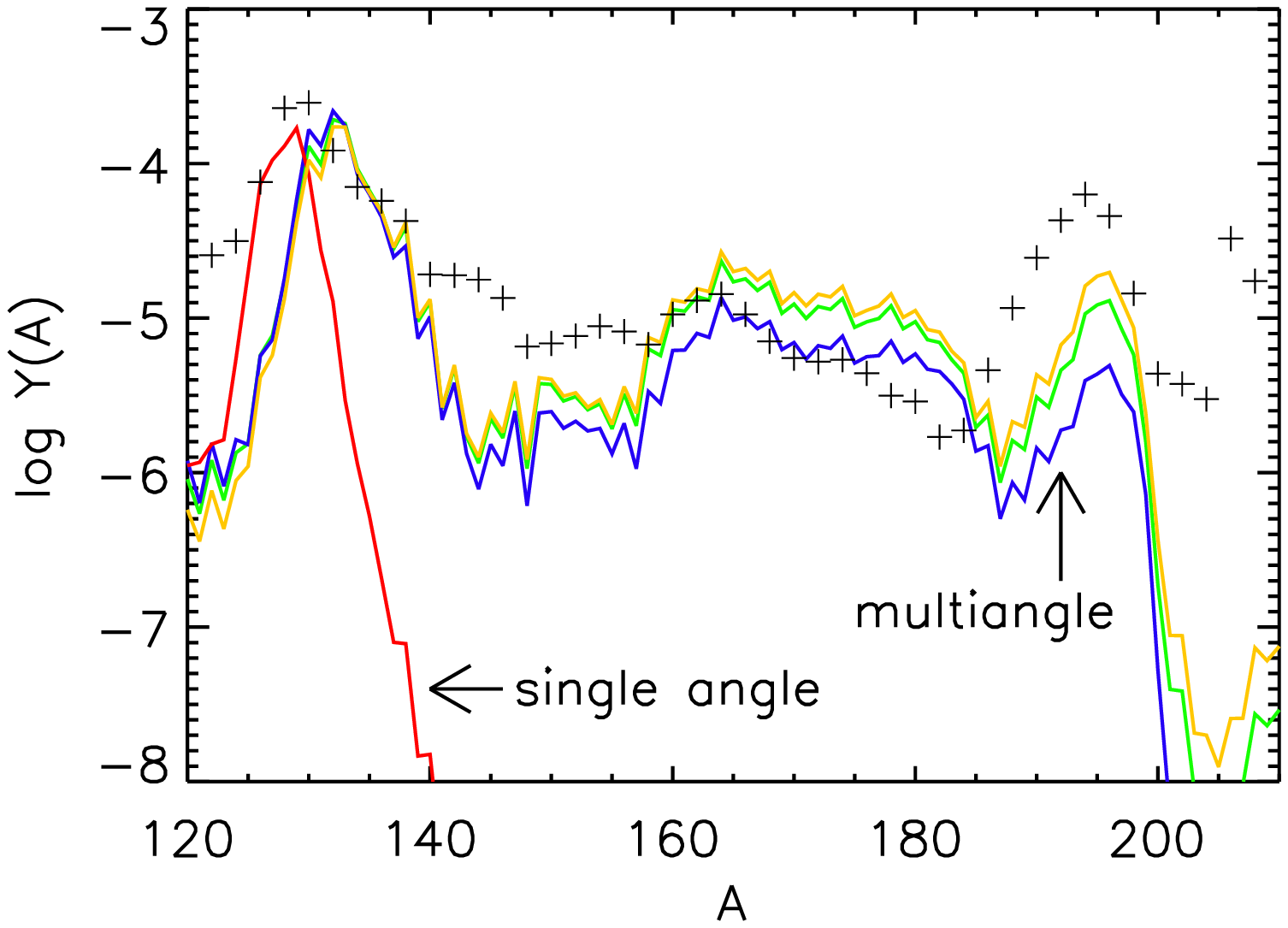,width=14cm}
          \caption{Shows final abundances $Y$ versus mass number $A$ for simulations with no neutrino
oscillations (green) and single-angle (red) and full multiangle (blue) oscillation calculations,
both assuming an inverted hierarchy.  Scaled solar abundances (crosses) and the results of a
simulation with neutrino interactions turned off at $T_{9}\sim 9$ (yellow) are shown for
comparison.  All four simulations use the early-type density profile with entropy $s/k=200$
and initial timescale $\tau=15$ ms.}
          \label{fig:YvAa}}

\FIGURE[p]{\epsfig{file=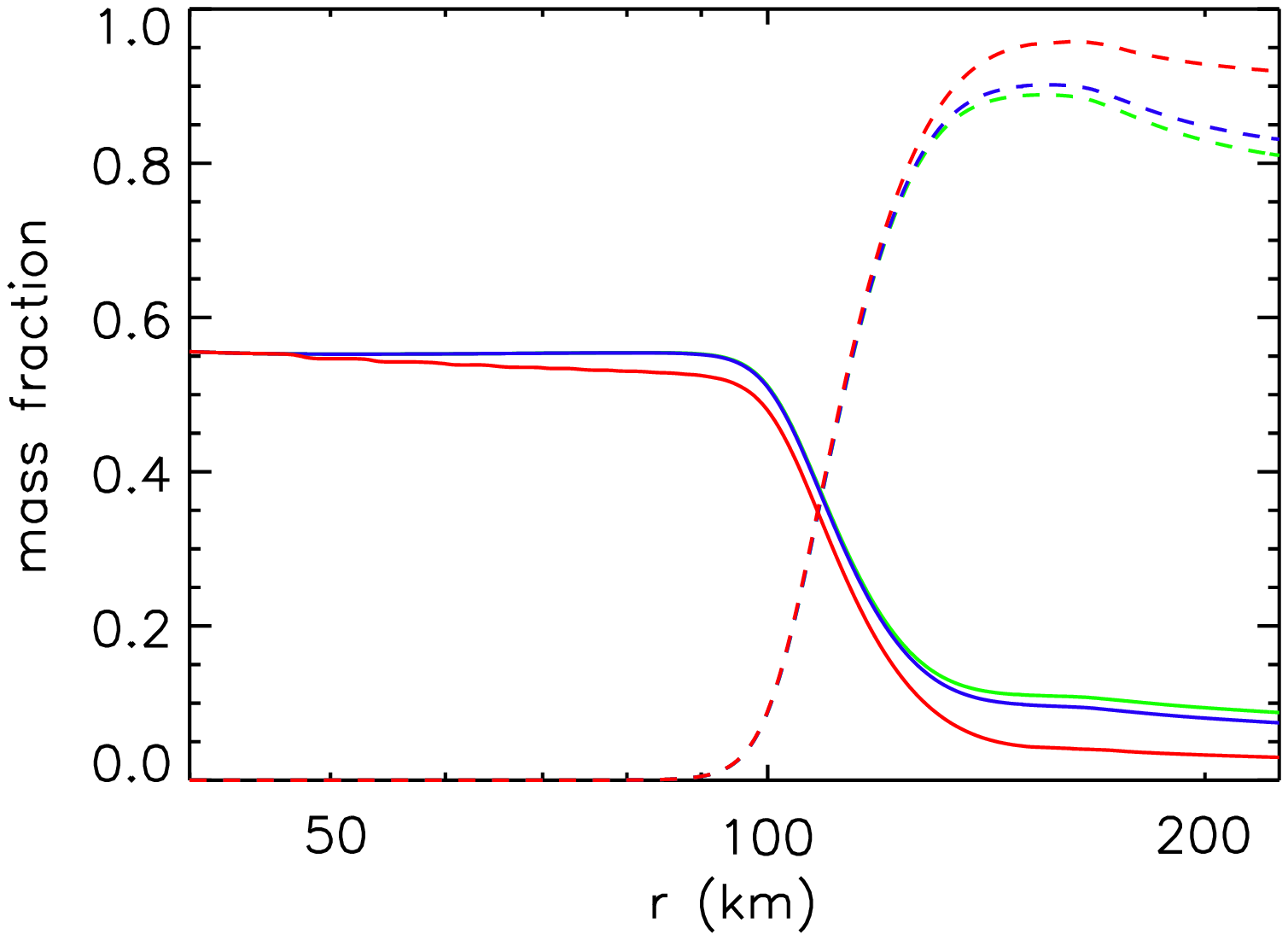,width=14cm}
          \caption{Shows mass fractions of neutrons (solid lines) and alpha particles (dashed lines) as
a function of radius $r$ in km for the three simulations from Fig.~\ref{fig:YvAa}.  In the single-angle
calculation, the flavor transformations occur early and influence the initial free neutron abundance, while in
the multiangle calculation the transformations occur as the alphas are assembling into seeds.}
          \label{fig:nuc}}

 We now compare the results discussed above to those with a calculation where the entropy have been altered (the
timescale has been correspondingly adjusted to ensure appropriate $r$-process conditions in the absence of
oscillations). Fig.~\ref{fig:YvAb} shows sample results from a set of simulations using a higher entropy
trajectory.  Here we use an early-type density profile with entropy $s/k=300$ and timescale $\tau=35$ ms, and
show results for simulations with no oscillations and with the full multiangle calculation for the oscillations. 
Note here the influence of the oscillations is stronger than in Fig.~\ref{fig:YvAa}; higher entropy pushes the
nucleosynthesis farther from the PNS, and so here the flavor transformations occur at a somewhat earlier stage
in the nucleosynthesis.  Furthermore the timescale is slower, which means that the material experiences a larger
neutrino fluence, i.e. more neutrinos and antineutrinos capture on nucleons, and this accentuates the effect of
oscillations.  Thus, the impact of flavor transformation is dependent on the thermodynamic and hydrodynamic
conditions.

We also consider the effect on the final abundance pattern due to the neutrino hierarchy.  The top panel of
Fig.~\ref{fig:YvAb} was produced with an inverted hierarchy, but in the bottom panel a normal hierarchy was
used.  As long as the $\mu$ neutrino luminosity $L_{\nu_{\mu}}$ is set to its $q=3.0$ Keil {\it et al} value, there is
little difference between the normal and inverted hierarchy cases.  However the results for the normal hierarchy
are quite sensitive to $L_{\nu_{\mu}}$.  As can be seen in the figure, a small decrease in the $\mu$ neutrino
luminosity completely shuts off the collective oscillations in the normal hierarchy. Thus, we find that in some
cases, the results will be insensitive to the hierarchy.  However, in the more general case, for a given
set of electron neutrino and anti-electron neutrino spectra, it is a combination of the hierarchy and the $\mu$
neutrino luminosity that determines the initial starting point of the flavor transformations as well as the course
that the oscillation takes.

\FIGURE[p]{\epsfig{file=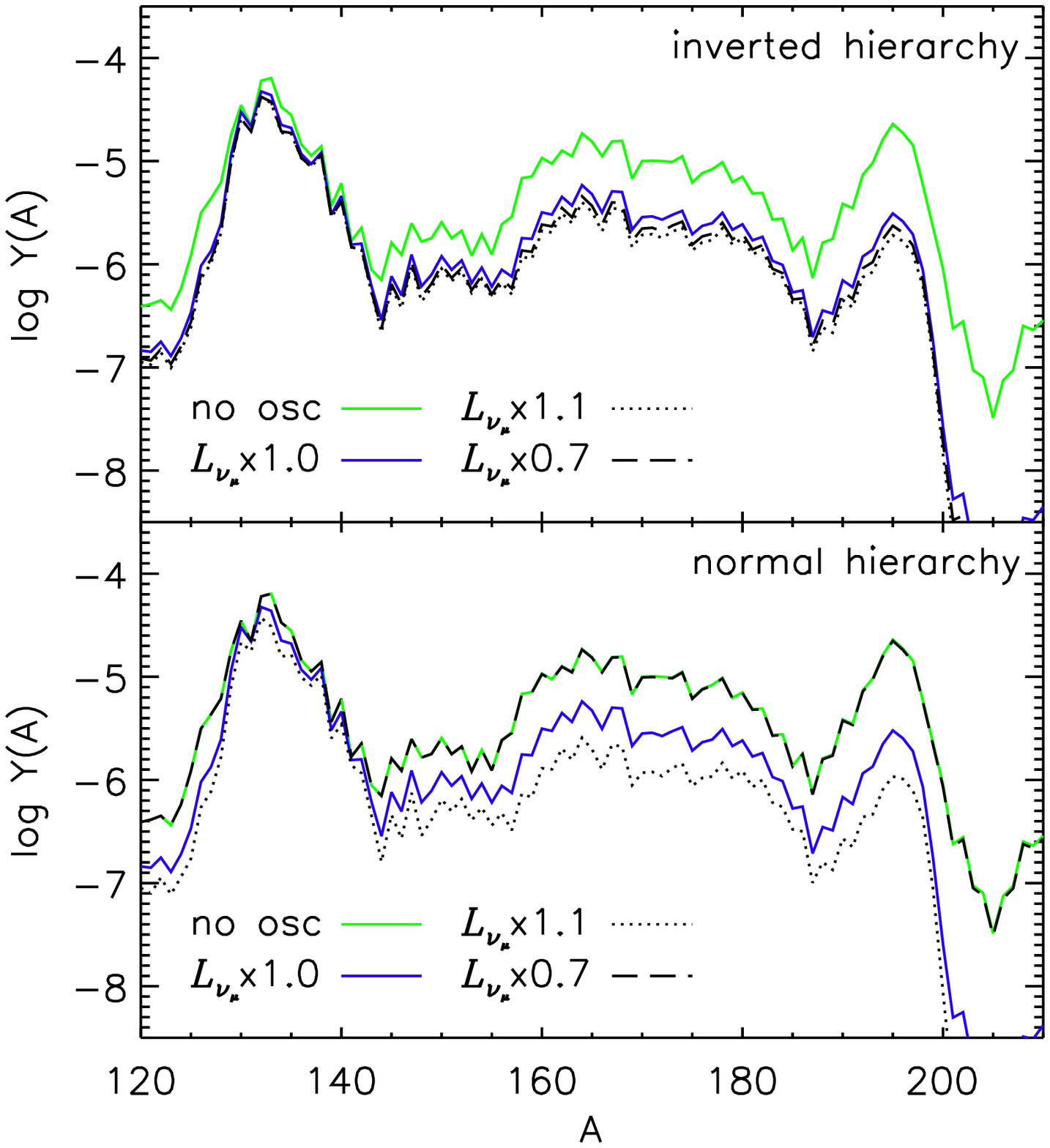,width=14cm}
          \caption{Shows final abundances $Y$ versus mass number $A$ for simulations with no neutrino
oscillations (green) and full multiangle oscillation calculations (blue) assuming a normal hierarchy
(bottom panel) and inverted hierarchy (top panel).  The results from simulations where the $\mu$
neutrino luminosity is multiplied by either a factor of 1.1 (dotted line) or 0.7 (dashed line) are also shown
for each case.  All simulations use the early-type density profile with entropy $s/k=300$ and initial timescale
$\tau=35$ ms.}
          \label{fig:YvAb}}

 Finally, we investigate the results for a different density profile, which has lower densities at a given
radius than the one we have used thus far. Fig.~\ref{fig:YvAc} shows results of simulations using a late-type
density profile, here with entropy $s/k=200$ and initial timescale 18 ms.  Nucleosynthesis moves inward, as
shown in Fig.~\ref{fig:nucc}, and so the flavor transformations occur at a later stage and the influence is
similar, but slightly reduced. Still, the size of the effect is similar to that which arises from
uncertainties in the nuclear physics inputs, as shown in, for example, Fig. 2 in \cite{Surman:2008ef} or Fig. 7
from \cite{Arcones:2010dz}.

Approximate radii for each stage of the nucleosynthesis are listed in Table~\ref{tab:r}.  Taken together with
the toy model presented Section 2, one can make a rough guess of the likely impact of collective flavor
transformation if one knows the point at which collective oscillation begin (which depends on the neutrino
spectra and luminosities).   However, we caution that we have found that significant deviations from this
rough guess can occur depending on the phase of the oscillation at each point in the evolution of the
composition of the material.

\FIGURE[p]{\epsfig{file=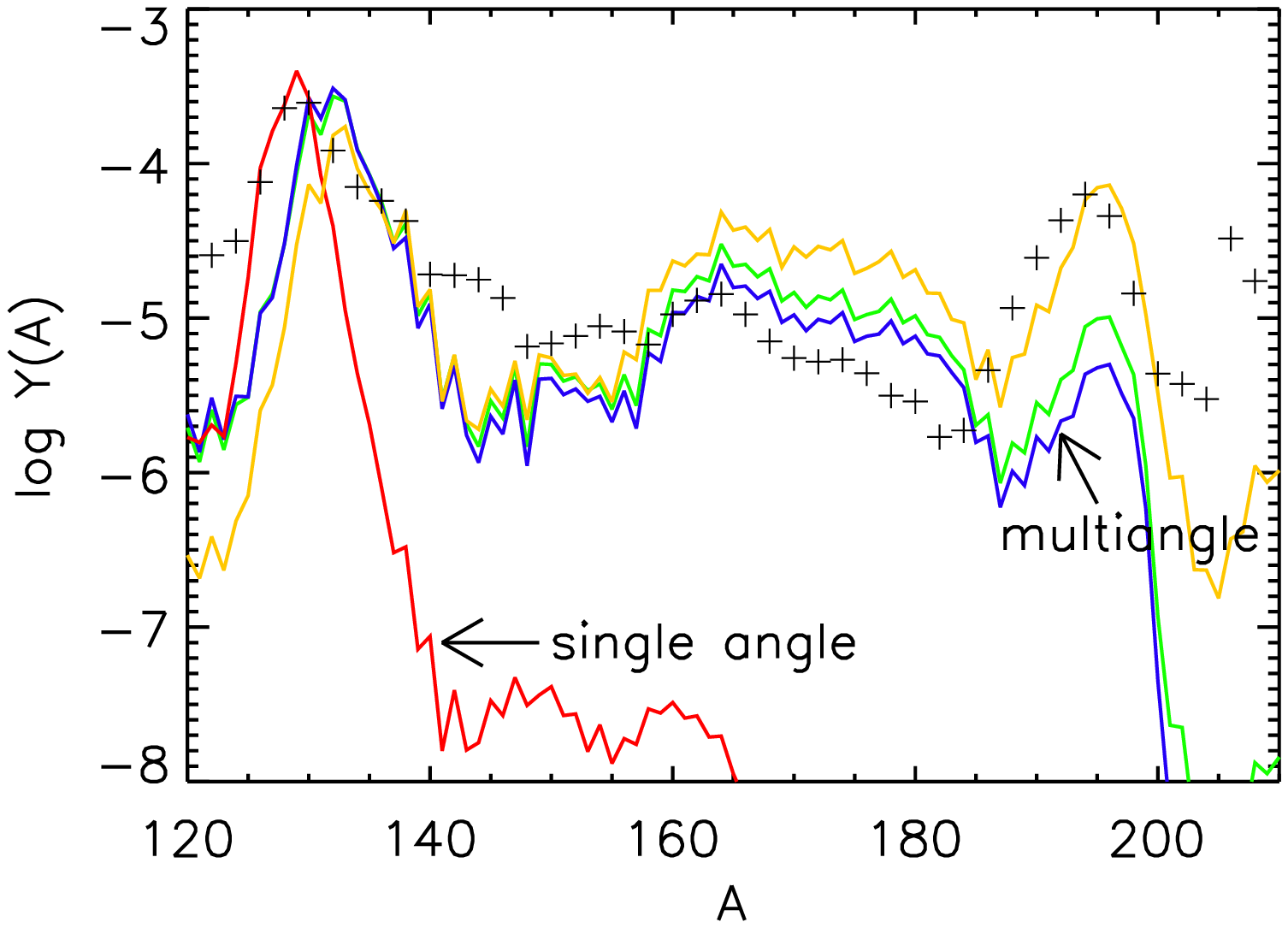,width=14cm}
          \caption{Shows final abundances $Y$ versus mass number $A$ for simulations with no neutrino
oscillations (green) and single-angle (red) and full multiangle (blue) oscillation calculations,
both assuming an inverted hierarchy.  Scaled solar abundances (crosses) and the results of a
simulation with neutrino interactions turned off at $T_{9}\sim 9$ (yellow) are shown for
comparison.  All four simulations use the late-type density profile with entropy $s/k=200$
and initial timescale $\tau=18$ ms.}
          \label{fig:YvAc}}

\FIGURE[p]{\epsfig{file=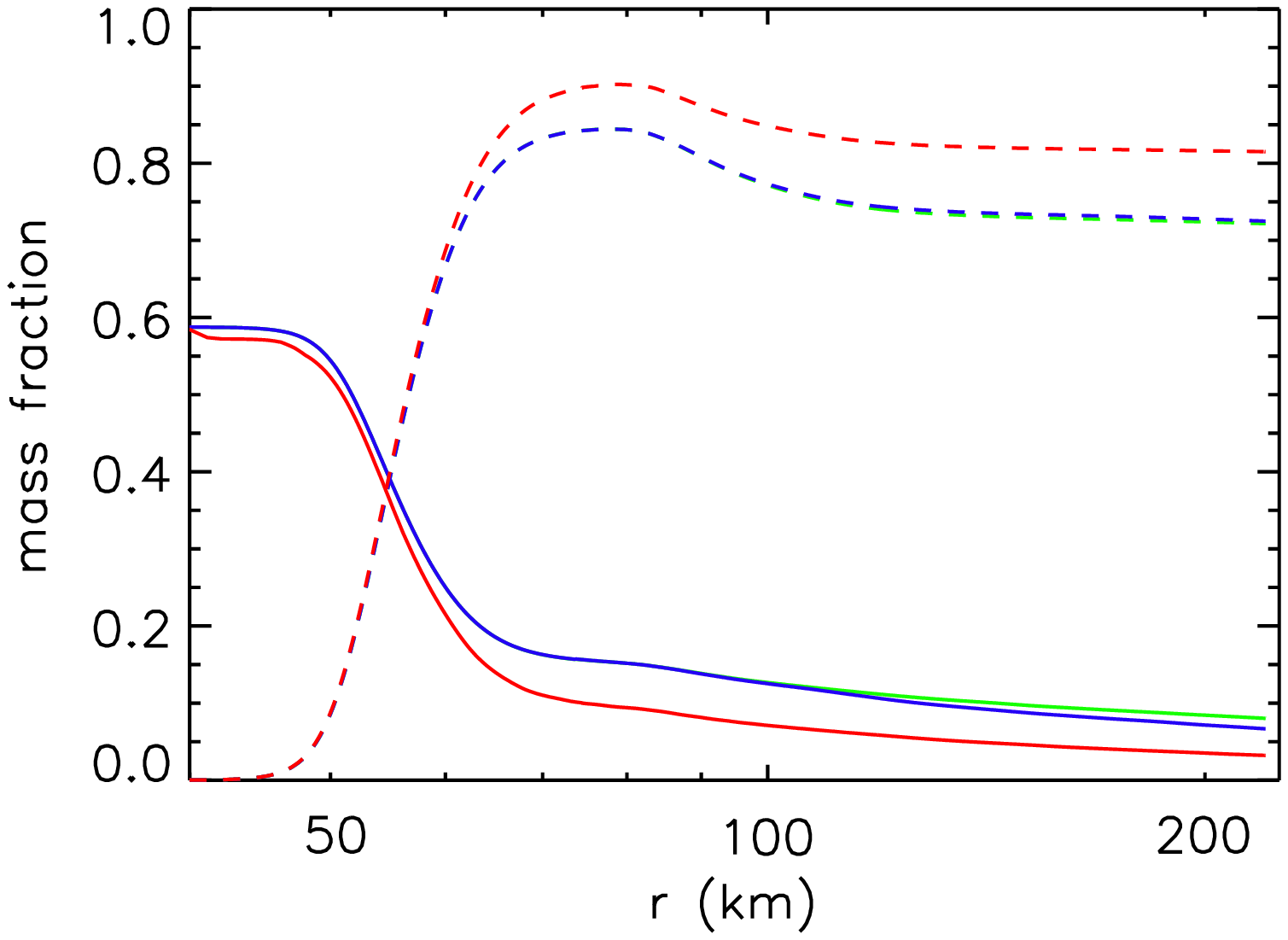,width=14cm}
          \caption{Shows mass fractions of neutrons (solid lines) and alpha particles (dashed lines) as
a function of radius $r$ in km for the three simulations from Fig.~\ref{fig:YvAc}.  The flavor
transformations occur at a significantly later stage of the nucleosynthesis than in the simulations
shown in Fig.~\ref{fig:nuc}.}
          \label{fig:nucc}}

\TABLE[p]{\begin{tabular}{| c | c | c | c | c |}
\hline
$s/k$ & density profile & $R_{\alpha}$ (km) & $R_{\mathrm{seed}}$ (km) & $R_{r-\mathrm{process}}$ (km) \\ \hline \hline
100 & early-type & 80 & 120 & 350 \\ \hline
200 & early-type & 100 & 170 & 440 \\ \hline
300 & early-type & 120 & 200 & 500 \\ \hline
100 & late-type & 40 & 60 & 175 \\ \hline
200 & late-type & 50 & 85 & 220 \\ \hline
300 & late-type & 60 & 100 & 250 \\ 
\hline
\end{tabular}
	\caption{Shows the approximate radii in km at which various stages of the nucleosynthesis begin for a range of
thermodynamic trajectories. We define $R_{\alpha}$ ($R_{\mathrm{seed}}$) as the radius at which the mass fraction of
alpha particles (seed nuclei) is 10\% of its maximum value, and $R_{r-\mathrm{process}}$ is the radius at which
$T_{9}\sim 2$.}
	\label{tab:r}}

\section{Conclusion}
\label{sec:conclusions}

We establish, using full three neutrino flavor multi-angle calculations, that the flavor transformation can
begin in regions sufficiently close to the center of a proto-neutron star that the outcome of element synthesis
in hot outflows is appreciably affected.  Further, the effect on the final abundance distribution is comparable
to that which stems from uncertainties in the nuclear conditions, e.g. \cite{Surman:2008ef,Arcones:2010dz}.
Neutrino collective oscillations act to enhance the role that neutrinos play in $r$-process nucleosynthesis, and
decreases the efficacy of the rapid neutron capture process. We find that the effect of collective flavor
transformation can be dramatically overestimated if the ``single-angle'' approximation is made.  An observable
effect on the abundance pattern can occur in both the normal and inverted hierarchy.  We find that the outcome
is dependent not only on the spectra and luminosity of the electron and antielectron neutrinos but also on the
$\mu$- and $\tau$-type neutrino spectra.

We stress again that the nucleosynthesis calculations are sensitive to how collective oscillations develop early on,
whereas neutrino signal in a terrestrial detector measures the final spectra after the oscillations cease. There are
known cases where the final spectra come out qualitatively similar in the single-angle and multiangle calculations, even
though at the intermediate stages the spectra are radically different \cite{Duan:2010bf}. In this sense, the
nucleosynthesis calculations place high demands on the accuracy of the oscillations calculations.

We continue the tradition of studying neutron rich outflows from proto-neutron stars that might occur in supernovae.  
While self-consistent models which produce a sufficiently neutron rich environment in these conditions have remained 
elusive, this site remains under consideration as a potential environment due to a variety of astrophysical
indicators that favor core-collapse supernovae as the origin of the ``main'' $r$-process elements.  If and when
a self consistent model of a hot outflow is produced, we have shown that these neutrino flavor transformation effects
will need to be included. In the scenarios we have considered, the flavor transformation effects make it more
difficult to produce $r$-process nuclei.  However, one must keep in mind that if new, more neutron rich possibilities are
discovered, the flavor transformation effects could potentially work in the direction
of improving the comparison with abundance data.

Our work improves on previous efforts to estimate the effect of the $r$-process in two important ways.  (1) We use three
flavor multi-angle calculations as opposed to ``single angle'' and/or two flavor calculations.  (2) We couple our three
flavor multi-angle calculation to a nuclear reaction network so that we include the effects of material composition. 
These are both essential for making reasonable quantitative estimates of the impact of flavor transformation.

There are a variety of types of elements that are made in the presence of a strong neutrino flux.  This type of
environment occurs in supernovae near proto-neutron stars, and in compact object mergers and gamma ray bursts
from disks around black holes, e.g.~\cite{Surman:2005kf,Surman:2008qf,Metzger:2007}.  In these strong neutrino
fluxes, as we have discussed, neutrinos can be expected to transform collectively. While it may be possible to
find some environments with hot outflows where the abundance yields are unaffected by neutrino flavor
transformation, this is not clear a priori.  The analysis presented here suggests that flavor transformation
must be considered in order to predict accurate abundance yields in environments with strong neutrino fluxes.

\acknowledgments

This work was supported in part by DOE contract DE-FG02-02ER41216 (GCM) and DE-FG05-05ER41398 (RS), by the DOE topical
collaboration,``Neutrinos and Nucleosynthesis in Hot and Dense Matter'' (HD, AF, and GCM), and by LANL LDRD program (HD
and AF).

\end{document}